\documentclass[aps,amsmath,amsfonts,11pt, preprint]{revtex4}
\usepackage{graphicx}
\usepackage{braket}
\usepackage{epstopdf}
\usepackage{epsfig}
\usepackage[english]{babel}
\usepackage{amsfonts}
\usepackage{physics}
\usepackage{amsmath, amssymb}
\usepackage{latexsym}
\usepackage{graphics,bm}
\usepackage{dcolumn}
\usepackage{rotating}
\usepackage{algorithm}
\usepackage{algpseudocode}
\usepackage{float}

\begin{document}
\title{Automorphism-Assisted QAOA: A Classical-Estimator Speedup for QAOA Simulation on Graphs with Non-Trivial Symmetry}

\author{Vaibhav N Prakash$^{1}$}

\affiliation{$^{1}$Department of Physics, Mahindra University, Hyderabad-500043, India}

\date{\today}

\begin{abstract}
We present Automorphism-Assisted QAOA (AA-QAOA), an observable-substitution method that accelerates the classical statevector simulation of QAOA on graphs whose automorphism group is non-trivial. For the unweighted MaxCut Hamiltonian $H_C = \tfrac{1}{2}\sum_{(i,j)\in E}(1-\sigma_z^{(i)}\sigma_z^{(j)})$, the cost expectation aggregation costs $\mathcal{O}(|E|\cdot 2^n)$ wall time on a statevector estimator. We show that on the Aut$(G)$-symmetric state prepared by the standard QAOA ansatz, replacing $H_C$ with an orbit-reduced observable $H_{\mathrm{red}}$, retaining one representative ZZ term per edge orbit weighted by orbit size, satisfies $\langle H_{\mathrm{red}}\rangle = \langle H_C\rangle$ exactly, dropping the aggregation cost to $\mathcal{O}(|\mathrm{orbits}|\cdot 2^n)$ with no change to the prepared state, optimization landscape, or approximation ratio. We benchmark AA-QAOA on tree-structured graphs up to 34 vertices and on six non-tree families, including the complete graph $K_n$, the star graph, and random 3-regular graphs, as controlled diagnostics that isolate the mechanism. The complete graph $K_{16}$ at $p=1$ shows an $8\times$ wall-time reduction despite spanning every qubit in the causal cone of its single representative edge, showing that the saving tracks orbit count rather than the qubit span of the representative reverse causal cones. A CPU/GPU control reproduces the speedup on both backends, confirming an architecture-independent, classical-aggregation origin. Approximation ratios satisfy $R_{\mathrm{red}}\approx R_{\mathrm{full}}$ throughout. We are explicit about scope: this is a speedup for classical-simulation pipelines, with no benefit on a quantum processor, where the full circuit executes regardless of the measured observable. The method is therefore of direct use to research groups running QAOA simulation at scale without QPU access.
\begin{description}
\item[Keywords]
QAOA classical simulation, statevector estimator, graph automorphism, edge equivalence classes, observable reduction, MaxCut.
\end{description}
\end{abstract}
\maketitle
\section{Introduction}
The Quantum Approximate Optimization Algorithm (QAOA), introduced by Farhi et al.~\cite{far2014}, has become a leading method for addressing NP-hard combinatorial optimization challenges on near-term quantum devices. As a hybrid quantum-classical algorithm, QAOA optimizes a parameterized quantum circuit through variational techniques, exploring solution spaces by alternating unitaries informed by problem-specific Hamiltonians. This approach presents a robust framework for efficiently solving complex optimization problems like MaxCut, where classical solutions are often computationally prohibitive due to the exponential search space associated with all potential graph cuts.

In the context of the unweighted MaxCut problem, QAOA constructs a Hamiltonian that encodes graph connectivity, with each term corresponding to an edge that signifies whether adjacent vertices belong to the same partition. Recent research has focused on enhancing the efficiency of QAOA, addressing the algorithm's computational demands and improving performance on large-scale instances~\cite{adapt-qaoa, shortcut-qaoa, qaoa2, fermionic-qaoa, scale-qaoa, mfoa, warm-qaoa, greedy-qaoa}. A primary bottleneck in QAOA lies in optimizing gate parameters, which scales significantly with the number of qubits even at minimal circuit depths (e.g., single Trotterized layer). Consequently, leveraging graph symmetries, especially automorphisms~\cite{classic-sym, obstacle-sym, quantum-sym, small-sym, error-sym}, has emerged as a promising strategy. Automorphisms are bijective mappings that preserve graph structure, enable the identification of equivalence classes among graph edges, thus reducing redundancy within the solution space. For instance,~\cite{exploit-qaoa} explores an automorphism-based simplification that calculates optimal gate parameters via the Reverse Causal Cone (RCC). This method identifies edge equivalence classes based on automorphism group generators $Aut(G)$, and computes the ground-state energy classically using tensor-networks for each of the RCC subgraph associated with a representative edge from each equivalence class. Automorphisms are computed using the highly efficient Nauty (No Automorphisms, Yes?) package, developed by McKay and Piperno~\cite{nauty}, which is widely used in complex graph structures. 

Building on these principles, we propose Automorphism-Assisted QAOA (AA-QAOA), a method that accelerates the classical simulation of QAOA by replacing the cost Hamiltonian $H_C$ with an orbit-reduced observable $H_{\mathrm{red}}$ that retains one representative ZZ term per edge orbit, weighted by orbit size. The full QAOA ansatz, built from $H_C$, is executed unchanged. Only the observable evaluated by the classical statevector estimator at every optimization step is reduced. Because the prepared state is invariant under Aut$(G)$, we show in Section~\ref{RH} that $\langle H_{\mathrm{red}}\rangle = \langle H_C\rangle$ exactly, so optimization against $H_{\mathrm{red}}$ converges to the same parameters and the same approximation ratio as optimization against $H_C$. Where the prior literature on automorphism exploitation in QAOA~\cite{exploit-qaoa, classic-sym} targets the reverse causal cone (RCC) of representative edges to argue for a circuit-level reduction, our framing is explicitly tied to the classical estimator: on a quantum processor the full circuit executes regardless of the measured observable, and the saving we report does not transfer to hardware. The intended use case is therefore the substantial body of QAOA research that runs entirely on classical statevector simulation, where the expectation aggregation $\sum_{(i,j)\in E} c_{ij}\langle Z_iZ_j\rangle$ at every optimizer iterate dominates the wall time and scales linearly in the number of measured terms. We provide an \textbf{empirical analysis} on tree-structured graphs up to 34 vertices and on six non-tree families that span the symmetry regime, including diagnostic controls that isolate the mechanism of the speedup as the orbit-count reduction rather than the qubit span of the representative RCCs.

The article is organized as follows. In Section~\ref{A0} we briefly review the unweighted MaxCut problem followed by a description of graph automorphisms in Section~\ref{A01}. In Section~\ref{section4} we provide a detailed algorithm to find edge equivalence classes which are used to reduce the number of terms in problem Hamiltonian discussed in Section~\ref{RH}. The results are provided and discussed in Section~\ref{RD} before concluding the article in Section~\ref{concl}.

\section{QAOA and the unweighted MaxCut problem}\label{A0}
The Quantum Approximate Optimization Algorithm (QAOA) is a hybrid quantum-classical algorithm designed to solve combinatorial optimization problems. The algorithm was introduced by Farhi et al.~\cite{far2014} and has since been widely studied for its potential to provide approximate solutions to NP-hard problems. QAOA is a variational quantum algorithm. It operates by constructing a parameterized quantum circuit with a sequence of alternating unitary operators that encode both the problem constraints and the exploration of the solution space.

QAOA starts by defining two key Hamiltonians: the problem Hamiltonian \( H_P \), which encodes the objective function of the problem, and the mixer Hamiltonian \( H_M \), which drives transitions between different states in the solution space. For a \( p \)-layer QAOA ansatz, the quantum state is constructed as:

\begin{equation}
|\psi(\vec{\beta}, \vec{\gamma})\rangle_p = \prod_{l=1}^{p} e^{-i \beta_l H_M} e^{-i \gamma_l H_C} |+\rangle^{\otimes n}
\end{equation}

where \( |+\rangle^{\otimes n} \) is the initial equal superposition state, and \( \vec{\beta} = (\beta_1, \beta_2, \dots, \beta_p) \), \( \vec{\gamma} = (\gamma_1, \gamma_2, \dots, \gamma_p) \) are the variational parameters to be optimized. The objective of QAOA is to minimize (or maximize) the expectation value of the problem Hamiltonian \( H_P \) with respect to the quantum state \( |\psi(\vec{\beta}, \vec{\gamma})\rangle_p \), i.e., the cost function is:

\begin{equation}
F_p(\vec{\beta}, \vec{\gamma}) = \langle \psi(\vec{\beta}, \vec{\gamma}) | H_P | \psi(\vec{\beta}, \vec{\gamma}) \rangle_p
\end{equation}

The parameters \( \vec{\beta} \) and \( \vec{\gamma} \) are iteratively updated by a classical optimizer to minimize this cost function, such that:

\begin{equation}
(\vec{\beta}^*, \vec{\gamma}^*) = \arg \min_{\vec{\beta}, \vec{\gamma}} F_p(\vec{\beta}, \vec{\gamma})
\end{equation}

In the case of the MaxCut problem, the problem Hamiltonian \( H_P \) is defined as:

\begin{equation}
H_P = \frac{1}{2} \sum_{(i,j) \in E} w_{ij} (1 - \sigma_z^{(i)} \sigma_z^{(j)})
\end{equation}

where \( w_{ij} \) is the weight of the edge between vertices \(i\) and \(j\), and \( \sigma_z^{(i)} \) is the Pauli-Z operator acting on qubit \(i\). In this article we only look at the unweighted MaxCut problem where,
\begin{equation}
    w_{ij} = \begin{cases}
        0, & \text{if no edge between vertices $i$ and $j$} \\
        1, & \text{if edge between $i$ and $j$}
    \end{cases}
\end{equation}
In this sense $w_{ij}$ is derived from the adjacency matrix of the corresponding graph $G$. The mixer Hamiltonian, independent of the problem, is typically defined as:

\begin{equation}
H_M = \sum_{i=1}^n \sigma_x^{(i)}
\end{equation}

where \( \sigma_x^{(i)} \) is the Pauli-X operator, which induces transitions between different states in the computational basis, ensuring exploration of the solution space. The overall quantum circuit consists of alternating layers of unitary operators corresponding to \(H_P\) and \(H_M\):

\begin{equation}
\begin{split}
    \mathcal{U}(\vec{\beta}, \vec{\gamma}) &= \prod_{l=1}^{p} e^{-i \beta_l H_M} e^{-i \gamma_l H_P}\\
    &= \prod_{l=1}^{p} \mathcal{U_M}(\beta) \mathcal{U_P}(\gamma)
\end{split}
\end{equation}

For each layer, the unitaries are parameterized by \(\beta_l\) and \(\gamma_l\), allowing the quantum state to evolve based on these parameters. The QAOA ansatz can be seen as a time-discretized version of adiabatic quantum computation, and for \( p \to \infty \), the algorithm converges to the exact solution of the problem by finding the ground state of \( H_P \). The optimization process involves measuring the expectation value of \( H_P \) repeatedly by executing the quantum circuit multiple times, also known as ``shots,'' and using these measurements to refine \( \vec{\beta} \) and \( \vec{\gamma} \) in subsequent iterations. The number of shots required for statistical accuracy scales as \( O(\epsilon^{-2})\ \)~\cite{vqerev}, where \( \epsilon \) is the desired precision. For larger systems, this number can grow exponentially, making QAOA's efficiency highly dependent on the quantum device and optimization method used. The QAOA algorithm is iterative, with each iteration involving both quantum circuit evaluations and classical optimization steps. For combinatorial optimization problems like MaxCut, QAOA shows promising results, with performance typically measured in terms of the approximation ratio:

\begin{equation}
R = \frac{F_p(\vec{\beta}^*, \vec{\gamma}^*)}{C_{\text{max}}}
\end{equation}

where \( C_{\text{max}} \) is the optimal cut value of the graph. Thus, QAOA provides an approximate solution to NP-hard problems by variationally optimizing the parameters of a quantum circuit.

\section{Graph Automorphisms}\label{A01}
In graph theory, an automorphism \( \phi : V \to V \) is a bijective mapping of the vertex set \( V \), where the adjacency relationships between vertices are preserved. Specifically, for any pair of vertices \( u, v \in V \), the automorphism \( \phi \) ensures that \( (\phi(u), \phi(v)) \in E \) if and only if \( (u, v) \in E \). This property implies that \( \phi \) maps the graph onto itself while maintaining its structure and connectivity, thereby revealing the inherent symmetries of the graph. The collection of all such automorphisms forms a group under composition, known as the automorphism group \( \text{Aut}(G) \). This group encapsulates all symmetries of the graph, including the identity automorphism, which leaves all vertices unchanged. For a label-independent binary objective function \( f : \{0,1\}^n \to \mathbb{R} \) defined on a graph \( G \), an automorphism \( \phi \) is considered a symmetry of \( f \) if it preserves the function's value. A label-independent function depends only on the structural properties of the graph rather than the specific labeling of its vertices. This property is common in unweighted graphs, where the function's value is determined solely by the graph's topology. One prominent example of such an optimization problem is the unweighted MaxCut problem, where we aim to maximize the cut size, independent of vertex labeling, by leveraging the symmetries within the graph.

In this work, we investigate symmetries of an objective function \( f \) defined over \( n \)-bit strings, where a symmetry is described by a permutation \( a \in S_{2^n} \). Such a permutation maps an input string \( x \) to a permuted version \( a(x) \) while preserving the function's value, i.e., \( f(x) = f(a(x)) \) $\forall$ \( x \in \{0, 1\}^n \). We focus on a specific class of symmetries called variable index permutations, which permute the positions of the variables in the bit strings, for instance. They are denoted by the symmetric group \( S_n \), acting on \( n \) variables. When transitioning to a quantum representation, these symmetries can be naturally captured using unitary operators. For any permutation \( a \in S_n \), there exists a corresponding unitary matrix \( A \) that performs the equivalent permutation on the quantum states. Formally, the unitary operator \( A \) is defined as:
\begin{equation}\label{A1}
    A = \sum_x |a(x)_1 \dots a(x)_n \rangle \langle x_1 \dots x_n |.
\end{equation}
This operator acts on a quantum state \( |x\rangle \), such that \( A|x\rangle = |a(x)\rangle \), permuting the basis states in accordance with the permutation \( a \). In the case where the objective function exhibits a symmetry \( a \in S_n \), the Hamiltonian \( H \), which represents the objective function in the quantum setting, must also respect this symmetry. This is guaranteed by the condition \( A^\dagger H A = H \), ensuring that the Hamiltonian remains invariant under the action of the unitary operator \( A \) corresponding to the symmetry.

Additionally, the symmetries of the problem manifest in both the initial state and the mixing Hamiltonian employed in the Quantum Approximate Optimization Algorithm (QAOA). The initial state \( |+\rangle \), often chosen as a uniform superposition over all bit strings, satisfies \( A|+\rangle = |+\rangle \) for any \( A \in S_{2^n} \), indicating that the state is invariant under the permutation symmetries. Furthermore, the mixing Hamiltonian \( B = \sum_{j=1}^n \sigma_x^{(j)} \), where \(\sigma_x^{(j)} \) is the Pauli-X operator on qubit \( j \), respects the symmetry as well. This is expressed by \( A^\dagger B A = B \), showing that the mixing Hamiltonian remains unchanged under the action of the symmetry operator \( A \).
If $\mathcal{U_P}(\gamma)=e^{-iH\gamma}$ and $\mathcal{U_M}(\beta)=e^{-iB\beta}$ implies, $\comm{A}{\mathcal{U_P}(\gamma)}=0=\comm{A}{\mathcal{U_M}(\beta)}$. We can use this to further evaluate the action of $A$ on a parameterized arbitrary state $\ket{\Vec{\beta},\Vec{\gamma}}_p$, after applying p-Trotterized layers on the initial superposition state $\ket{+}$,
\begin{equation}\label{A2}
\begin{split}
     A\ket{\Vec{\beta},\Vec{\gamma}}_p &= A\ \mathcal{U_M}(\beta_p)\mathcal{U_P}(\gamma_p)\cdots \mathcal{U_M}(\beta_1)\mathcal{U_P}(\gamma_1)\ket{+}\\
     &=\mathcal{U_M}(\beta_p)\mathcal{U_P}(\gamma_p)\cdots \mathcal{U_M}(\beta_1)\mathcal{U_P}(\gamma_1)\ A\ket{+}\\
     &=\ket{\Vec{\beta},\Vec{\gamma}}_p.
\end{split}
\end{equation}
Consider a subset of qubit indices, $S_1$ and $S_2$ such that $a(S_1)=S_2$.  If observables $\prod_{j\in S_1}\sigma_z^{(j)}$ and $\prod_{j\in S_2}\sigma_z^{(j)}$ are connected by permutation $a\in S_n : a(S_1)=S_2$ then $A^\dagger\prod_{j\in S_1}\sigma_z^{(j)} A = \prod_{j\in S_2}\sigma_z^{(j)}$. This can be straightforwardly understood as the \textbf{transformation $A^\dagger\sigma_z^{(j)}A$ changing the set of qubits on which the product of Pauli-Z operators can act on iff the set of qubits are connected by a permutation $a\in S_n$}. Using Eq.~\ref{A2} we can further decipher,
\begin{equation}\label{A3}
    \begin{split}
        \bra{\Vec{\beta},\Vec{\gamma}}\prod_{j\in S_1}\sigma_z^{(j)}\ket{\Vec{\beta},\Vec{\gamma}}_p &=\bra{\Vec{\beta},\Vec{\gamma}}A^\dagger\prod_{j\in S_1}\sigma_z^{(j)} A\ket{\Vec{\beta},\Vec{\gamma}}_p \\
        &=\bra{\Vec{\beta},\Vec{\gamma}}\prod_{j\in S_2}\sigma_z^j\ket{\Vec{\beta},\Vec{\gamma}}_p.
    \end{split}
\end{equation}
To reduce the number of terms in the problem Hamiltonian for the unweighted MaxCut problem, we identify edge equivalence classes (details provided in Section \ref{section4}). For each edge equivalence class \( \tilde{S}_m \), the subset of qubit indices corresponding to an edge, denoted by \( s_m \in \tilde{S}_m \), is connected to other edge subsets in the same class through $Aut(G)$. From Eq.~\ref{A3} it becomes clear that the measurements of all the edge terms in the problem Hamiltonian can be replaced by weighted measurements of only certain terms representing each edge equivalence class, the weights being the degeneracy (class length) of each class.
 
\section{Automorphism-Based Graph Processing}\label{section4}

The Nauty package (No Automorphisms, Yes?) is a widely-used tool for computing automorphism groups of graphs. In the context of unweighted graphs, it efficiently computes generators for the automorphism group to determine edge equivalence classes. These equivalence classes categorize edges based on the symmetries present in the graph, specifically how automorphisms can map one edge to another.

\textbf{Algorithm 1} describes the procedure for computing edge equivalence classes using the generators of the automorphism group. The process starts with the initialization of graph structures and a generator array to store automorphisms, which are stored as permutations. As automorphisms are identified, they are applied to the edges of the graph. For each edge \((u, v)\), the generators transform it into a permuted edge \((u_{\text{perm}}, v_{\text{perm}})\), and these transformed edges are output for further processing.

The next stage of the algorithm groups edges into equivalence classes by iterating over each edge and applying the automorphism generators. If the permuted edge matches an existing equivalence class, the edge is assigned to that class. If no matching class is found, a new equivalence class is created. \textbf{The equivalence between edges in the same class is determined by the fact that they can be mapped onto each other via a generator of the automorphism group}, meaning they share identical structural roles in the graph's symmetry. Thus, two edges belong to the same equivalence class if there exists a generator from the automorphism group that transforms one edge into the other. Automorphisms preserve the adjacency relations between vertices, classifying edges based on this symmetry.

\begin{algorithm}[H]
\caption{Automorphism-Based Edge Orbit Computation}
\label{alg:automorphism}
\begin{algorithmic}[1]

\Function{EDGE\_EQUIVALENCE}{$G$}
    \State \textbf{Input:} $G$ -- a graph with vertex set $V$ and edge set $E$.
    \State \textbf{Output:} $C$ -- list of edge-equivalence classes $\{\tilde{S}_1, \tilde{S}_2, \ldots\}$.
    \State Build sparse adjacency structure $sg$ from $G$.
    \State \textbf{Call} \texttt{sparsenauty($sg$)} $\to$ generators $\{g_1, \ldots, g_k\}$ of $\mathrm{Aut}(G)$ (stored via callback).
    \State Collect all edges $(u,v)$ with $u < v$ into sorted array \textit{edges}; mark all unclassified.
    \State $num\_classes \leftarrow 0$
    \For{each unclassified edge $(u, v)$ in \textit{edges}}
        \State $num\_classes \mathrel{+}= 1$; create class $\tilde{S}_{num\_classes} \leftarrow \{(u,v)\}$; mark $(u,v)$ classified.
        \State Initialize queue $Q \leftarrow [(u, v)]$.
        \While{$Q$ is not empty}
            \State Dequeue $(u', v')$ from $Q$.
            \For{$i = 1$ \textbf{to} $k$}
                \State $(u'', v'') \leftarrow (g_i(u'),\; g_i(v'))$
                \If{$(u'', v'')$ is an edge \textbf{and} unclassified}
                    \State Add $(u'', v'')$ to $\tilde{S}_{num\_classes}$; mark classified; enqueue $(u'', v'')$.
                \EndIf
            \EndFor
        \EndWhile
    \EndFor
    \State \Return $C = \{\tilde{S}_1, \ldots, \tilde{S}_{num\_classes}\}$
\EndFunction
\end{algorithmic}
\end{algorithm}

\section{Reducing the Hamiltonian}\label{RH}
The central correctness anchor of AA-QAOA is the identity
\begin{equation}\label{eq:exact}
\langle \psi(\vec{\beta},\vec{\gamma}) | H_{\mathrm{red}} | \psi(\vec{\beta},\vec{\gamma}) \rangle_p \;=\; \langle \psi(\vec{\beta},\vec{\gamma}) | H_C | \psi(\vec{\beta},\vec{\gamma}) \rangle_p,
\end{equation}
which holds exactly, not approximately, on the Aut$(G)$-invariant state prepared by the standard QAOA ansatz of Eq.~\ref{A1}. The derivation follows directly from Eq.~\ref{A2}, which establishes that $A|\psi(\vec{\beta},\vec{\gamma})\rangle_p = |\psi(\vec{\beta},\vec{\gamma})\rangle_p$ for every $A$ corresponding to a graph automorphism, together with Eq.~\ref{A3}, which shows that the expectation value of a representative ZZ term equals that of every other ZZ term in its edge orbit. Writing $H_C = \sum_{(i,j)\in E}\tfrac{1}{2}(1-Z_iZ_j)$ and grouping by edge orbit $\tilde{S}_m$ with representative $(i_m, j_m)$ and orbit size $|\tilde{S}_m|$,
\begin{equation}
H_{\mathrm{red}} \;=\; \tfrac{1}{2}\sum_{m=1}^{|\mathcal{O}_E|} |\tilde{S}_m|\,\bigl(1 - Z_{i_m}Z_{j_m}\bigr),
\end{equation}
where $|\mathcal{O}_E|$ is the number of edge orbits. Eq.~\ref{eq:exact} is the algebraic content that makes the substitution lossless. The optimal parameters $(\vec{\beta}^*,\vec{\gamma}^*)$ are identical under both observables, and so are the approximation ratios.

The Ising Hamiltonian can comprise a large number of terms, especially for graphs with many edges. Each term corresponds to a pair of vertices connected by an edge, and the total number of terms increases linearly with $|E|$. The orbit substitution replaces the linear-in-$|E|$ sum with a sum over $|\mathcal{O}_E|$ representatives, which can be substantially smaller. The speedup of AA-QAOA arises from the consequent reduction of the classical estimator's expectation aggregation cost, derived explicitly in Section~\ref{est-cost}.

For unweighted graphs, the original QUBO matrix for the MaxCut problem contains off-diagonal elements set to either zero or one, indicating the presence or absence of an edge, while diagonal elements remain zero, as self-edges are disallowed. We modify this QUBO matrix by setting all off-diagonal elements to zero, except for those corresponding to the representative edges from each equivalence class. The value of each retained off-diagonal element is then set to the size of its respective equivalence class. As a result, the original QUBO matrix for the MaxCut problem is replaced by a modified version, where non-zero off-diagonal elements represent the representative edges and their values reflect the size of the corresponding edge equivalence classes.
The next step involves transforming the modified QUBO matrix into the corresponding Ising Hamiltonian, $H_{red}$, using the \texttt{QuadraticProgram()} function from the \texttt{qiskit-optimization} module. The Ising Hamiltonian $H_{red}$, derived from the modified QUBO matrix, contains fewer terms compared to the original Hamiltonian $H_C$ for the MaxCut problem. We then use the reduced Ising Hamiltonian $H_{red}$ as our observable, measured at the output of the QAOA circuit constructed from the full Ising Hamiltonian $H_C$.

We retain the discussion of the \textbf{Reverse Causal Cone (RCC)} below for two reasons. First, the RCC framework is the standard language in which graph-symmetry exploitation in QAOA has historically been discussed~\cite{exploit-qaoa}, and our placement of AA-QAOA in that landscape requires the construct. Second, the RCC of the representative edges gives a natural structural descriptor, denoted $\eta_p$ in later sections, that we use to characterize families of graphs in our experiments. We are explicit, however, that the RCC is not the source of the speedup we report. The classical estimator runs the full circuit at every iterate and the cost saving derives from the aggregation step downstream, as we make precise in Section~\ref{est-cost}.

\subsection{Reverse Causal Cone}
Originally proposed in the original QAOA article by Farhi et al.~\cite{far2014}, a $\textit{Reverse Causal Cone (RCC)}$ is a selection of qubits interacting in $p$ Trotterized layers with the measured qubits $i$ and/or $j$ via one and two-qubit gates. Consider the computation of expectation value of $\sigma_z^{(i)}\sigma_z^{(j)}$ operator in the eigenstate of the $\sigma_x$ operator (initial state in vanilla QAOA):
\begin{equation}\label{rcc1}
    \begin{split}
        \expval{\sigma_z^{(i)}\sigma_z^{(j)}}{\psi(\Vec{\beta},\Vec{\gamma})} &= 
        \bra{+} \mathcal{U_P}^\dagger(\gamma_1) \mathcal{U_M}^\dagger(\beta_1) \cdots 
        \mathcal{U_P}^\dagger(\gamma_p) \mathcal{U_M}^\dagger(\beta_p) \sigma_z^{(i)} \sigma_z^{(j)} \\
        &\quad \mathcal{U_M}(\beta_p) \mathcal{U_P}(\gamma_p) \cdots 
        \mathcal{U_P}(\gamma_1) \mathcal{U_M}(\beta_1) \ket{+} \\
        &= 
        \bra{+} \mathcal{U_P}^\dagger(\gamma_1) \mathcal{U_M}^\dagger(\beta_1) \cdots 
        \mathcal{U_P}^\dagger(\gamma_p) \mathcal{U_M}^\dagger(\beta_p) \sigma_z^{(i)} \mathcal{U_M}(\beta_p) \\
        &\quad \mathcal{U_M}^\dagger(\beta_p)\sigma_z^{(j)} \mathcal{U_M}(\beta_p) 
        \mathcal{U_P}(\gamma_p) \cdots \mathcal{U_P}(\gamma_1) \mathcal{U_M}(\beta_1) \ket{+}.
    \end{split}
\end{equation}

for $p$ Trotterized layers. In the second step we have split the identity operator into a product of unitaries but it also suggests that the one qubit gates in the last layer of QAOA do not affect the correlation between qubits $i$ and $j$. Using a placeholder operator $\mathcal{O}(i,j)$ and considering $p=2$, without loss of generality, we can re-write Eq.~\ref{rcc1} as:
\begin{equation}\label{rcc2}
    \begin{aligned}
        \expval{\sigma_z^{(i)}\sigma_z^{(j)}}{\psi(\vec{\beta}, \vec{\gamma})} 
        &= \bra{+} \mathcal{U_P}^\dagger(\gamma_1) \mathcal{U_M}^\dagger(\beta_1) 
            \mathcal{U_P}^\dagger(\gamma_2) \mathcal{O}(i,j) \mathcal{U_P}(\gamma_2) 
            \mathcal{U_P}(\gamma_1) \mathcal{U_M}(\beta_1) \ket{+} \\
        &= \bra{+} \mathcal{U_P}^\dagger(\gamma_1) \mathcal{U_M}^\dagger(\beta_1) 
            \left( \prod_{k,l \in N(\{i,j\})} \mathcal{U_P}^\dagger(\gamma_2)_{(k,l)}
            \mathcal{O}(i,j) \prod_{k,l \in N(\{i,j\})} \mathcal{U_P}(\gamma_2)_{(k,l)} \right) \\
        &\quad \mathcal{U_M}^\dagger(\beta_p) \sigma_z^{(j)} \mathcal{U_M}(\beta_p) 
            \mathcal{U_P}(\gamma_p) \mathcal{U_P}(\gamma_1) \mathcal{U_M}(\beta_1) \ket{+}.
    \end{aligned}
\end{equation}

where we have broken $\mathcal{U_P}(\gamma_2)$ as the product of two-qubit gates corresponding to the shared edges in problem hamiltonian with either qubit $i$ or qubit $j$. The product is over only those qubits in the second layer which directly connect to qubit $i$ or $j$ (neighbours of $(i,j)$) via two-qubit gates. All the remaining two-qubit gates will commute through and give identity. If $\{k\}$ is a set of all such qubits which are neighbours of $i$ or $j$ then $N(\{i,j\})=\{i,j\}\cup \{k\}$. Thus Eq.~\ref{rcc2} can be written as,
\begin{equation}\label{rcc3}
    \begin{split}
        \expval{\sigma_z^{(i)}\sigma_z^{(j)}}{\psi(\Vec{\beta},\Vec{ \gamma})} &= 
        \bra{+} \mathcal{U_P}^\dagger(\gamma_1) \mathcal{U_M}^\dagger(\beta_1) 
        \mathcal{O}(N(\{i,j\}))
        \mathcal{U_P}(\gamma_1) \mathcal{U_M}(\beta_1) \ket{+}.
    \end{split}
\end{equation}
Progressing further with the first layer will give us the placeholder operator which contains qubits connected via two-qubit gates to qubits $\in N(\{i,j\})$. This creates a network of qubits connected to qubits whose correlations we are trying to find. This network or subgraph is the \textit{RCC} for qubits $i$ and $j$. In other words, the \textit{RCC} for a qubit pair $(i,j)$ in QAOA refers to the set of qubits that influence the measurement of qubits $i$ and $j$ through two-qubit gates in $p$ Trotterized layers of QAOA. The expectation value of $\sigma_z^{(i)}\sigma_z^{(j)}$ operator can be re-written as:
\begin{equation}\label{rcc4}
    \begin{split}
        \expval{\sigma_z^{(i)}\sigma_z^{(j)}}{\psi(\Vec{\beta},\Vec{ \gamma})} &= 
        \bra{RCC_{subgraph}} \sigma_z^{(i)}\sigma_z^{(j)}\ket{RCC_{subgraph}}.
    \end{split}
\end{equation}
Correlations between different qubits will generate different RCC subgraphs. The question is, how many Trotterized layers, $p$, in QAOA circuit derived from the full Ising Hamiltonian $H_{Full}$, will contain the complete RCC structure of all the pairs of qubits $(i,j)$ in the reduced Hamiltonian $H_{Red}$. This would determine $p$ for any problem instance which involves reduced Hamiltonians implemented via graph automorphism.
\subsection{Structural descriptor: combined-RCC coverage}\label{rcc-coverage}
\begin{figure}[h!]
    \centering
    \includegraphics[width=0.7\linewidth]{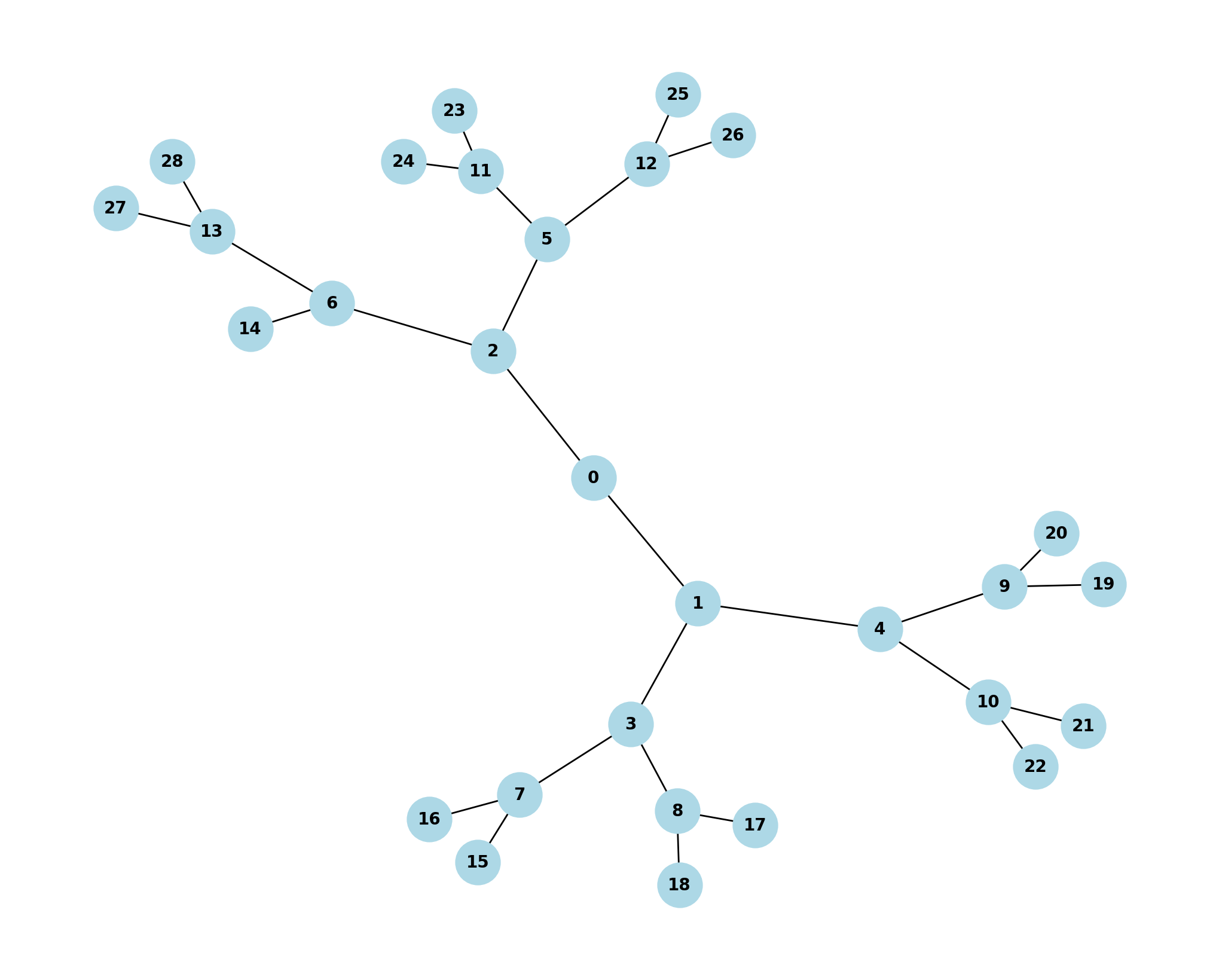}
    \caption{(Color Online) Binary Tree graph with 29 vertices and 28 edges, $G(29,28)$.}
    \label{fig1}
\end{figure}

For the binary tree $G(29,28)$ shown in Figure~\ref{fig1}, the 12 edge orbits give a combined RCC at $p=1$ that covers $\{0,1,2,3,4,5,6,7,8,11,12,13,14,15,16,23,24,27,28\}$ and leaves $\{9,10,17,18,19,20,21,22,25,26\}$ uncovered. At $p=2$, the RCCs expand by one hop and the union covers the full vertex set. The structural descriptor that quantifies this coverage is
\begin{equation}\label{eta}
    \eta_p \;=\; 1 \;-\; \frac{\big|\bigcup_{e\in\mathcal{R}}\mathrm{RCC}_p(e)\big|}{|E|},
\end{equation}
the fraction of graph edges that lie outside the combined RCC of the representative edges $\mathcal{R}$ at depth $p$. The descriptor is bounded by construction, $\eta_p\in[0,1]$, with $\eta_p\to 1$ when the representative RCCs are localized and $\eta_p = 0$ when their union spans every edge. We use $\eta_p$ in Sections~\ref{RD} and~\ref{diag} as a structural characterization of the graph family but, anticipating the diagnostic controls in Section~\ref{diag}, $\eta_p$ does not predict the wall-time speedup we measure: the complete graph $K_n$ has $\eta_p$ rising with $n$ and yet a single representative edge whose RCC spans every qubit, while the random 3-regular family has $\eta_p \approx 0$ at all sizes. The actual predictor of the speedup is the orbit count $|\mathcal{O}_E|$, as derived in Section~\ref{est-cost}.

\subsection{The classical estimator cost and the orbit reduction}\label{est-cost}
We now make precise the source of the AA-QAOA wall-time saving. The MaxCut cost Hamiltonian $H_C = \tfrac{1}{2}\sum_{(i,j)\in E}(1 - Z_iZ_j)$ is diagonal in the computational basis, so all its ZZ terms commute and are jointly diagonalisable. On a quantum processor this means the entire expectation $\langle H_C\rangle$ can be estimated from a single batch of computational-basis shots, with the per-term aggregation a trivial post-processing step on classical bitstrings. The number of terms carries no measurement-budget cost on hardware, and the AA-QAOA reduction gives no benefit there.

The classical statevector simulator is different. The standard QAOA simulation pipeline, of which Qiskit's \texttt{EstimatorV2} with \texttt{AerSimulator} as backend is representative, computes $\langle H_C\rangle$ by first evolving the statevector $|\psi(\vec{\beta},\vec{\gamma})\rangle$ through the QAOA circuit and then aggregating
\begin{equation}\label{eq:agg}
\langle H_C \rangle \;=\; \sum_{(i,j) \in E} c_{ij}\,\langle \psi | Z_iZ_j | \psi \rangle,
\end{equation}
where each $\langle Z_iZ_j\rangle$ requires an $\mathcal{O}(2^n)$ pass over the statevector amplitudes. The wall time of the aggregation step is therefore $\mathcal{O}(|E|\cdot 2^n)$ per optimizer iterate. For a COBYLA run with $T$ iterates the cumulative aggregation cost is $\mathcal{O}(T\cdot |E|\cdot 2^n)$, which dominates the cost of the circuit evolution itself when $|E|$ is comparable to or larger than the depth of the QAOA circuit, the typical regime at low $p$.

When the AA-QAOA substitution is applied, Eq.~\ref{eq:agg} is replaced by
\begin{equation}
\langle H_{\mathrm{red}}\rangle \;=\; \sum_{m=1}^{|\mathcal{O}_E|} |\tilde{S}_m|\,\langle \psi | Z_{i_m}Z_{j_m} | \psi\rangle,
\end{equation}
with $|\mathcal{O}_E|$ representative terms in place of $|E|$. The wall time of the aggregation step drops to $\mathcal{O}(|\mathcal{O}_E|\cdot 2^n)$, giving an expected speedup factor of $|E|/|\mathcal{O}_E|$. The statevector $|\psi\rangle$ itself is unchanged, since the QAOA ansatz is built from the full $H_C$. The exactness of Eq.~\ref{eq:exact} ensures that the optimizer sees the same scalar objective at every iterate.

Two consequences follow. First, the predictor of the AA-QAOA speedup is the orbit ratio $|E|/|\mathcal{O}_E|$, not $\eta_p$ and not the qubit span of the representative RCCs. We verify this prediction directly in Section~\ref{diag} using the complete graph $K_n$, which has $|\mathcal{O}_E|=1$ and exhibits the largest measured speedup, and the random 3-regular family, which has $|\mathcal{O}_E|=|E|$ asymptotically and exhibits no speedup. Second, the saving is intrinsic to classical aggregation and does not transfer to a QPU run of the same circuit, where the full set of ZZ observables is co-measured at the cost of a single shot batch. The AA-QAOA method is, in this sense, a classical-simulation accelerator for graphs with non-trivial automorphism, of direct use to research groups whose QAOA work runs entirely on classical statevector simulators.

\subsection{The AA-QAOA pipeline}\label{AA-QAOA}
Based on the preceding derivation, the AA-QAOA simulation pipeline replaces the cost-Hamiltonian observable measured at every classical optimizer iterate, while leaving the QAOA ansatz built from the full $H_C$ untouched. The steps are as follows.
\begin{itemize}

    \item \textbf{Graph Definition:} 
    We begin by defining the graph \( G = (V, E) \), where \( V \) is the set of vertices and \( E \) is the set of edges.

    \item \textbf{Finding Edge Equivalence Classes Using Automorphisms:} 
    Next, we compute the automorphism group of graph \( G \) to identify edge equivalence classes using the generators of the group (Algorithm~\ref{alg:automorphism}).

    \item \textbf{Selecting a Representative Edge from Each Equivalence Class:} 
    From each edge equivalence class, we randomly choose one representative edge. This reduces the number of edges we need to consider, as we only focus on the representatives for the construction of the QUBO matrix.

    \item \textbf{QUBO Matrix Initialization:} 
    After selecting the representative edges, we initialize a QUBO matrix of size \( V \times V \), where \( V \) is the number of vertices in the graph. This matrix will serve as a modified MaxCut formulation.

    \item \textbf{Populating the QUBO Matrix:} 
    In this step, the QUBO matrix is populated. For each representative edge, the corresponding matrix element is assigned a non-zero value equal to the size of the equivalence class to which the edge belongs. All other matrix elements are set to zero. This modification of the standard MaxCut QUBO ensures that only the representative edges contribute to the optimization problem.

    \item \textbf{Conversion to Ising Hamiltonian:} 
    The modified QUBO matrix is then converted into an Ising Hamiltonian using standard quadratic programming techniques. This Hamiltonian is optimized by the QAOA to approximate the solution to the MaxCut problem.

    \item \textbf{QAOA Ansatz Construction:} 
    After constructing the modified Ising Hamiltonian, the QAOA ansatz is built using the original MaxCut Ising Hamiltonian, which includes all terms and edges. The QAOA ansatz involves alternating between two unitaries, one derived from the Ising Hamiltonian and another from a mixing Hamiltonian. The alternating sequence of these unitaries is parameterized by the angles \( \gamma \) and \( \beta \), which are optimized during the algorithm.

    \item \textbf{Optimization:} 
    The final step involves optimizing the variational parameters \( \gamma \) and \( \beta \) to minimize the expectation value of the Ising Hamiltonian. While the QAOA ansatz is constructed from the full Hamiltonian, only terms in the modified Ising Hamiltonian are measured. The optimization is performed classically, and the result is an approximate solution to the MaxCut problem on the graph \( G \).

    \item \textbf{Obtaining the Optimal Solution:} 
    The optimized parameters \( \gamma^*, \beta^* \) are used to generate a final quantum state. By measuring this state, we obtain a bitstring that corresponds to the approximate solution of the MaxCut problem. This bitstring is the result of the QAOA process and represents the partition of the vertices of \( G \) that maximizes the number of edges between the two partitions.

\end{itemize}

\textbf{Use of AI in this work.} The author used Claude (Anthropic, Sonnet 5) for assistance deriving and verifying the mathematical formalism of Sections~\ref{A01}--\ref{RH}, for analysis of the diagnostic-control experiments of Section~\ref{diag} used to identify the mechanism underlying the reported speedup and to rule out a candidate qubit-span explanation, for drafting assistance in the Results and Discussion, and for copy-editing of grammar, spelling, and internal consistency. All AI-assisted analysis, derivations, and content were independently verified against the underlying code and data, and critically reviewed, by the author, who takes full accountability for the manuscript's scientific claims.

\section{Results and Discussion}\label{RD}
We employed the Quantum Approximate Optimization Algorithm with $p=1$ layer to solve MaxCut instances on tree-structured graphs, with node counts of up to 34. Two types of tree graphs were considered: binary trees and balanced trees, both generated using the NetworkX Python library. A key distinction between these graphs is that binary trees can exhibit uneven branch depths, particularly when the number of leaf nodes, $n$, is not a perfect power of two, resulting in an unbalanced structure. This asymmetry affects the number of edge equivalence classes in binary and balanced trees. The number of edge equivalence classes corresponds to a reduction in the number of terms in the Ising Hamiltonian, as outlined in Section \ref{AA-QAOA}. Tables \ref{tab1} and \ref{tab2} present the number of edge equivalence classes and the corresponding terms in the Ising Hamiltonian for MaxCut instances on binary and balanced trees, respectively. For comparison, we also provide the total number of terms in the full Hamiltonian, prior to any reductions.
\begin{table}[h!]
 \centering
 \begin{tabular}{|c| c| c| c| c| c|}
 \hline
 G(V,E) & Edge Equiv. classes & $H_{reduced}$ & $H_{full}$ & $\eta_{p=1}$ & $\eta_{p=2}$\\ [0.5ex]
 \hline\hline
 (5,4) & 3 & 7 & 9 & 0.00 & 0.00\\
 \hline
 (10,9) & 7 & 15 & 19 & 0.00 & 0.00 \\
 \hline
 (15,14) & 3 & 9 & 29 & 0.38 & 0.11 \\
 \hline
 (20,19) & 11 & 24 & 39 & 0.10 & 0.00\\
 \hline
 (25,24) & 11 & 23 & 49 & 0.22 & 0.06\\
 \hline
 (30,29) & 13 & 30 & 59 & 0.22 & 0.05\\
 \hline
 (31,30) & 4 & 11 & 61 & 0.57 & 0.30\\
 \hline
 (34,33)  & 16 & 35 & 67 & 0.19 & 0.05\\ [0.5ex]
 \hline
 \end{tabular}
 \caption{Binary Tree graph instances up to 34 vertices, reporting the number of edge orbits $|\mathcal{O}_E|$, the number of terms in $H_{\mathrm{red}}$ and in the full Hamiltonian $H_C$, and the structural descriptor $\eta_p$ (Eq.~\ref{eta}) at $p=1$ and $p=2$. The $\eta_p$ values are means over random representative-edge selection, with $\eta_2 \le \eta_1$ by construction. The classical-estimator wall-time speedup tracks $|E|/|\mathcal{O}_E|$ rather than $\eta_p$, as confirmed by the diagnostic controls in Section~\ref{diag}.}
 \label{tab1}
\end{table}

\begin{table}[h!]
 \centering
 \begin{tabular}{|c| c| c| c| c| c|}
 \hline
 G(V,E) & Edge Equiv. classes & $H_{reduced}$ & $H_{full}$ & $\eta_{p=1}$ & $\eta_{p=2}$ \\ [0.5ex]
 \hline\hline
 (7,6) & 2 & 5 & 13 & 0.17 & 0.00 \\
 \hline
 (13,12) & 2 & 6 & 25 & 0.34 & 0.00\\
 \hline
 (15,14) & 3 & 7 & 29 & 0.38 & 0.11\\
 \hline
 (31,30) & 4 & 11 & 61 & 0.57 & 0.30\\ [0.5ex]
 \hline
 \end{tabular}
 \caption{Balanced Tree graph instances up to 31 vertices, reporting the number of edge orbits $|\mathcal{O}_E|$, the number of terms in $H_{\mathrm{red}}$ and in the full Hamiltonian $H_C$, and the structural descriptor $\eta_p$ (Eq.~\ref{eta}) at $p=1$ and $p=2$. The $\eta_p$ values are means over random representative-edge selection, with $\eta_2 \le \eta_1$ by construction.}
 \label{tab2}
\end{table}
Tables~\ref{tab1} and~\ref{tab2} report the edge-orbit count $|\mathcal{O}_E|$ alongside the term counts of $H_{\mathrm{red}}$ and $H_C$, and the structural descriptor $\eta_p$. For these bounded-degree tree instances, fewer edge orbits yield a smaller $H_{\mathrm{red}}$ and, by Section~\ref{est-cost}, a larger expected aggregation speedup. This is most pronounced for the $G(15,14)$ and $G(31,30)$ binary trees, where the symmetry around the root node produces a larger automorphism group than their neighbouring instances $G(10,9)$ and $G(30,29)$. The two smallest instances $G(5,4)$ and $G(10,9)$ have orbit ratios $|E|/|\mathcal{O}_E|$ close to one, and the saving they exhibit in Section~\ref{RD} is correspondingly small. Note that $\eta_p$ is reported in the same tables for structural completeness, but its role as a predictor is limited to bounded-degree graphs and breaks down on the dense and the low-symmetry families analyzed in Section~\ref{diag}.

\begin{figure}[h!]
    \centering
    \includegraphics[width=1 \linewidth]{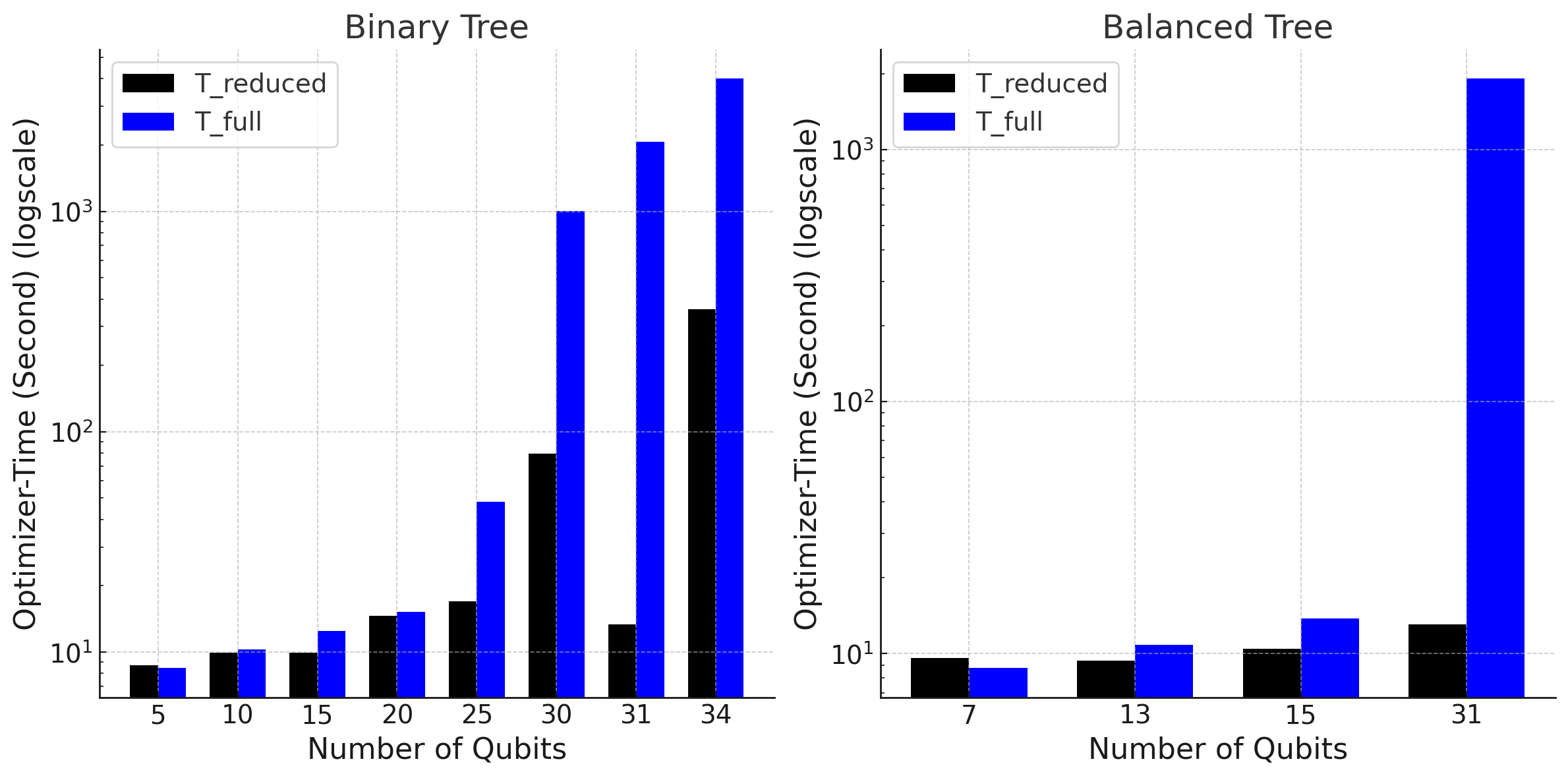}
    \caption{(Color Online) COBYLA optimizer-time bar plots for Binary and Balanced Tree graphs. Blue, Black bar denotes time taken for finding optimal parameters by measuring terms in full and reduced problem Hamiltonian respectively}
    \label{fig2}
\end{figure}

We conducted Python simulations on TPU-v2 cores, utilizing a system with 96 vCPUs and 350 GB of memory, provided through Google Colab. Although TPU acceleration for matrix multiplication was not utilized in this setup, the availability of a high number of vCPUs and substantial memory resources enabled the efficient execution of the simulations. The significance of this computational capacity is reflected in the results presented in Tables~\ref{tab3} and~\ref{tab4}. To compute the minimum eigenvalue, we employed the \texttt{EstimatorV2} class from \texttt{qiskit-ibm-runtime} module with \texttt{AerSimulator()} as backend. The circuit gate rotation parameters, $\beta$ and $\gamma$, were optimized using the classical COBYLA optimizer. Figure \ref{fig2} illustrates the optimizer-time required for each graph $G(V,E)$ in both binary and balanced tree structures to achieve the minimum eigenvalue.\\
Subsequently, we sampled the full quantum circuit using the optimal parameters obtained from the optimization process. The approximation ratios for the MaxCut problem were calculated for both the full Hamiltonian and the reduced Hamiltonian, with the latter leveraging graph automorphisms to reduce complexity. For MaxCut instances on tree-structured graphs, where no cycles exist, the exact optimal cut value $C_{max}$ equals the total number of edges in the graph. Tables \ref{tab3} and \ref{tab4} provide a detailed comparison of the optimizer-time, approximation ratio, and maximum system memory used for each instance.
\begin{table}[h!]
 \centering
 \resizebox{\textwidth}{!}{
 \begin{tabular}{|c|c|c|c|c|c|c|} 
 \hline
 G(V,E) & $T_{red}$(Sec) & $T_{full}$(Sec) & $R_{red}$ & $R_{full}$ & Peak Mem (Red) (GB) & Peak Mem (Full) (GB) \\ [0.5ex]
 \hline\hline
 (5,4)  & 8.72   & 8.45    & 1     & 1     & 5.2   & 5.2   \\ 
 \hline
 (10,9) & 9.89  & 10.25   & 1     & 1     & 5.2   & 5.2   \\
 \hline
 (15,14) & 9.90  & 12.50   & 1     & 1     & 5.3   & 5.3   \\
 \hline
 (20,19) & 14.56 & 15.27   & 0.84  & 0.84  & 5.3   & 5.3   \\
 \hline
 (25,24) & 16.97 & 48.25   & 0.83  & 0.83  & 5.4   & 5.8   \\
 \hline
 (30,29) & 79.55 & 1004.32 & 0.76  & 0.76  & 6.4   & 21.5  \\
 \hline
 (31,30) & 13.34 & 2067.36 & 0.80  & 0.80  & 5.4   & 37.4  \\
 \hline
 (34,33) & 360.56 & $>$3600 & 0.76  & 0.76  & 21.7  & 262.5 \\ 
 \hline
 \end{tabular}
 }
 \caption{Binary Tree Results}
 \label{tab3}
\end{table}

\begin{table}[h!]
 \centering
 \resizebox{\textwidth}{!}{
 \begin{tabular}{|c|c|c|c|c|c|c|} 
 \hline
 G(V,E) & $T_{red}$(Sec) & $T_{full}$(Sec) & $R_{red}$ & $R_{full}$ & Peak Mem (Red) (GB) & Peak Mem (Full) (GB) \\[0.5ex]
 \hline\hline
 (7,6)  & 9.63   & 8.75    & 1     & 1     & 5.2   & 5.2   \\ 
 \hline
 (13,12) & 9.38  & 10.82   & 1     & 1     & 5.2   & 5.2   \\
 \hline
 (15,14) & 10.44 & 13.78   & 0.93  & 0.93  & 5.2   & 5.3   \\
 \hline
 (31,30) & 13.02 & 1907.91 & 0.83  & 0.80  & 5.2   & 37.5  \\
 \hline
 \end{tabular}
 }
 \caption{Balanced Tree Results}
 \label{tab4}
\end{table}

The results presented in Tables~\ref{tab3} and~\ref{tab4} highlight the wall-time saving of the AA-QAOA observable substitution applied to classical statevector simulation of QAOA on tree-structured MaxCut instances. The reduction in optimizer time grows with the graph size, as expected from the aggregation argument of Section~\ref{est-cost}, since the per-iterate aggregation cost is $\mathcal{O}(|E|\cdot 2^n)$ under $H_C$ and $\mathcal{O}(|\mathcal{O}_E|\cdot 2^n)$ under $H_{\mathrm{red}}$, with both the $|E|$-versus-$|\mathcal{O}_E|$ ratio and the $2^n$ factor growing with graph size. The full-Hamiltonian run on the 34-qubit instance exceeded 3600 seconds on Colab and was terminated, against 360.56 seconds under $H_{\mathrm{red}}$, a reduction of more than $90\%$. Peak memory tracks the same pattern, with the reduced run consistently below the full run and substantially so on the larger instances, where the aggregation step dominates. The approximation ratios $R_{\mathrm{red}}$ and $R_{\mathrm{full}}$ are essentially identical in every row, consistent with Eq.~\ref{eq:exact}. The $G(15,14)$ and $G(31,30)$ binary trees stand out for the magnitude of their saving among the binary instances, traceable to their balanced-tree-like symmetry, which collapses orbit count well below that of their immediate neighbours $G(10,9)$ and $G(30,29)$. The mild excursion $R_{\mathrm{red}}=0.83$ versus $R_{\mathrm{full}}=0.80$ for $G(31,30)$ in Table~\ref{tab4} is an artifact of COBYLA optimizer variance at 31 qubits rather than a genuine improvement from the reduced observable, since the two observables share the optimization landscape by Eq.~\ref{eq:exact}.

All simulations in this section use $p=1$. Section~\ref{diag} reports a non-tree set of instances at both $p=1$ and $p=2$ to confirm that the wall-time saving persists at deeper circuits and is not specific to a single layer. The observation that $p=1$ already yields large savings on the binary trees follows from the orbit reduction of $H_C$ being depth-independent. The orbit structure of the graph is a property of the graph, not of the QAOA circuit, so the aggregation-cost ratio $|E|/|\mathcal{O}_E|$ is invariant under choice of $p$. What changes with $p$ is the $2^n$ prefactor (constant in $n$ here) and the optimizer iterate count, but the proportional saving from the aggregation step is preserved.

A first diagnostic instance is the star graph. With a single edge orbit, $H_{\mathrm{red}}$ retains one ZZ term while $H_C$ retains $n-1$. Naively, the aggregation cost ratio predicts a $(n-1)\times$ saving. In practice, at the moderate qubit counts of the original tree paper, the absolute aggregation time is small and overhead dominates. Table~\ref{tab5} reports the original measurements at $n=28$ and $n=29$, both of which sit near tie, an instance noise band rather than a refutation of the orbit-count prediction. The same star family at $n=16$ in the controlled study of Section~\ref{diag} (Table~\ref{tab:controls}) crosses out of the overhead band and exhibits a $3\times$ wall-time saving consistent with the aggregation argument.

\begin{table}[h!]
 \centering
 \resizebox{0.45\textwidth}{!}{
 \begin{tabular}{|c|c|c|c|c|c|c|} 
 \hline
 G(V,E) & $T_{red}$(Sec) & $T_{full}$(Sec) & $R_{red}$ & $R_{full}$ \\ 
 \hline\hline
 (28,27) & 79.16  & 83.94   & 0.963     & 1\\
 \hline
 (29,28)  & 187.26   & 190.16    & 0.964     & 0.964  \\ 
 \hline
 \end{tabular}
 }
 \caption{Table depicting optimizer-time for star-structured graph with 28 and 29 vertices.}
 \label{tab5}
\end{table}

The optimizer times for $G(28,27)$ and $G(29,28)$ are close to tie for both observables in Table~\ref{tab5}. With a single edge orbit and $n-1$ ZZ terms in $H_C$, the aggregation-cost ratio predicts a $(n-1)\times$ saving, but the measured absolute aggregation times at these qubit counts are dominated by COBYLA, transpilation, and Estimator setup overhead, and the orbit reduction sits inside that noise band. The slight excursion $R_{\mathrm{red}}=0.963$ versus $R_{\mathrm{full}}=1.0$ at $G(28,27)$ is COBYLA variance rather than a landscape perturbation, since the two observables share the landscape by Eq.~\ref{eq:exact}. Section~\ref{diag} reports a controlled star-family run from a separate codebase that crosses out of the overhead band at $n=16$ and shows the predicted saving.

\subsection{Diagnostic controls and the mechanism of the speedup}\label{diag}
The aggregation argument of Section~\ref{est-cost} predicts that the AA-QAOA wall-time saving is set by the orbit ratio $|E|/|\mathcal{O}_E|$, not by the qubit span of the representative RCCs and not by the choice of CPU or GPU backend. We test these predictions on three controlled axes designed to dissociate the candidate mechanisms. The first axis varies graph symmetry from edge-transitive (single orbit) to generically asymmetric (no automorphism beyond identity), to test whether the saving scales with orbit reduction. The second axis varies graph density from sparse star to dense complete, with the complete graph carrying a single edge orbit and a representative RCC that spans every qubit, to test whether qubit-span coincidence eliminates the saving. The third axis varies the simulation backend between CPU and GPU, to test architecture independence. Across the three axes, the empirical pattern is unambiguous.

\subsubsection{Structural descriptors across graph families}
\begin{figure}[htbp]
\centering
\includegraphics[width=\linewidth]{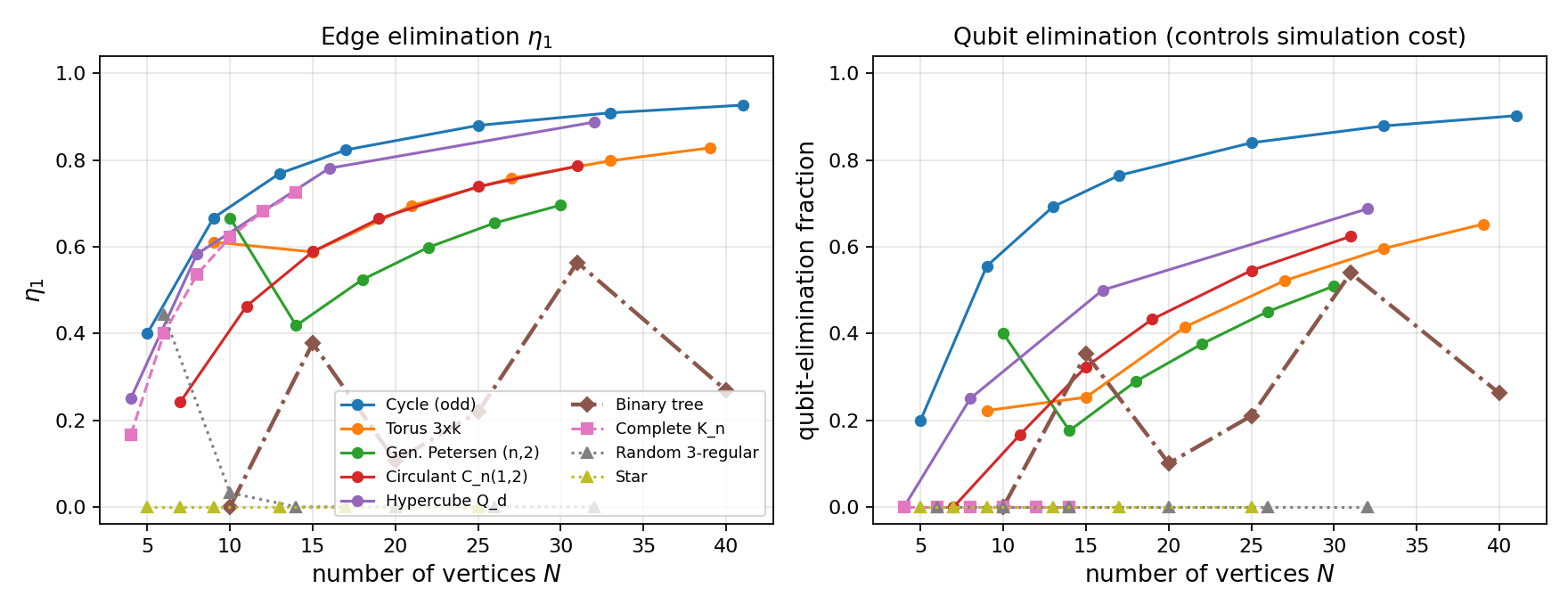}
\caption{(Color Online) Structural descriptors across eight graph families. Edge-elimination $\eta_1$ (left) and qubit-elimination fraction (right) versus number of vertices $N$, averaged over random representative selection. Bounded-degree edge-transitive families (solid) rise toward unity in both panels. The complete graph (dashed) rises in $\eta_1$ but is identically zero in qubit-elimination. The star and random 3-regular families (dotted) remain near zero in both. The dissociation between the two panels on $K_n$ is the structural test of whether the AA-QAOA saving tracks edges or qubits, resolved by the wall-time measurements in Figure~\ref{fig:speedup}.}
\label{fig:scaling}
\end{figure}

Figure~\ref{fig:scaling} reports the structural descriptor $\eta_1$ together with the related qubit-elimination fraction, both as functions of the number of vertices $N$ across eight graph families. For bounded-degree edge-transitive families (cycle, $3\times k$ torus, generalized Petersen, circulant, hypercube), the descriptors rise together toward unity as $N$ grows, since the representative RCC stays local while $|E|$ grows. For the complete graph $K_n$, $\eta_1$ rises with $N$ while qubit-elimination remains identically zero, because the single representative RCC spans every qubit at any size. For the star graph, both descriptors remain identically zero, because the central vertex forces the representative RCC to span the entire edge set. For random 3-regular graphs at $N\geq 14$, the automorphism group collapses to the identity, so $|\mathcal{O}_E|=|E|$ and both descriptors are zero. The dissociation between $\eta_1$ and qubit-elimination on $K_n$ is the structural signal that they index different things, and we use the wall-time measurements below to determine which of the two, if either, tracks the actual saving.

\subsubsection{Wall-time controls: complete graphs, stars, and random regular}
\begin{figure}[htbp]
\centering
\includegraphics[width=\linewidth]{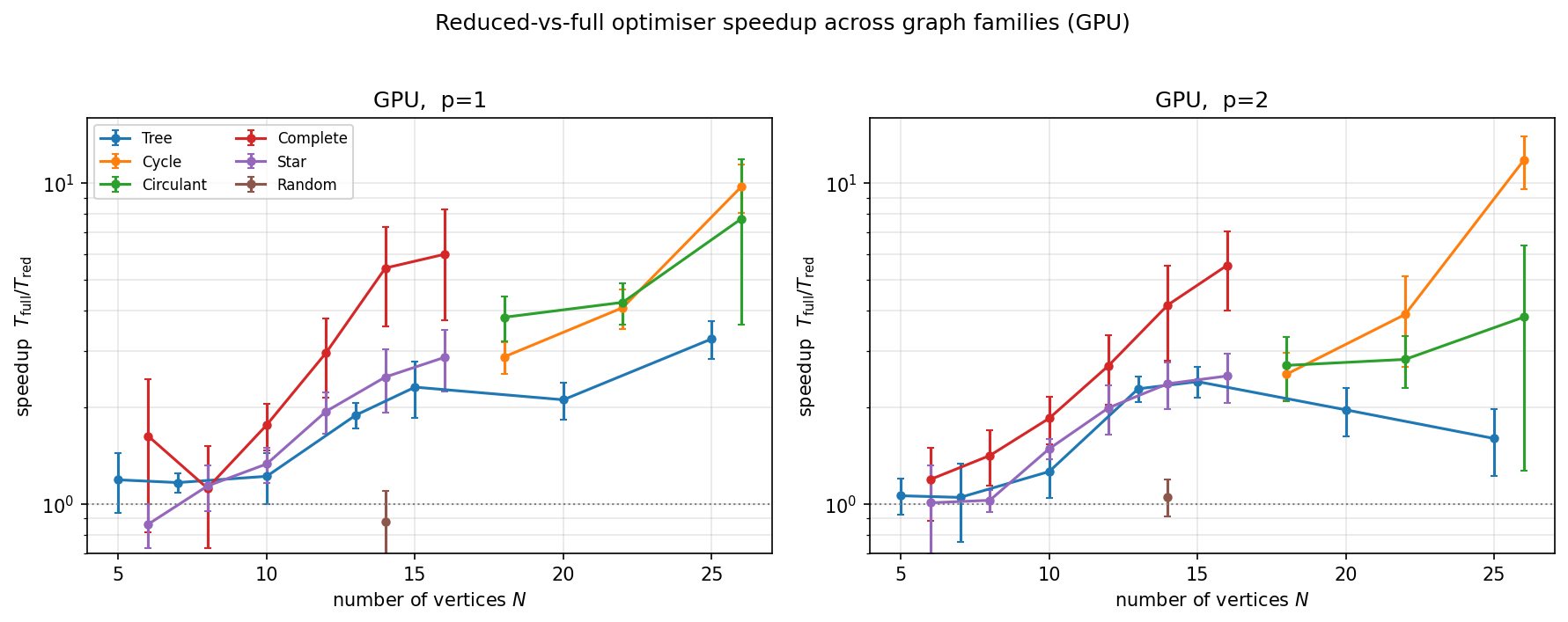}
\caption{(Color Online) Wall-time saving versus $N$ on the diagnostic-control families, CPU backend. The complete and star families show a saving that grows substantially with $N$, while the random 3-regular family is flat at unity. Both control families have representative RCCs that span every qubit, so the saving cannot be attributed to qubit-span elimination. The data are consistent with the orbit-count prediction of Section~\ref{est-cost}.}
\label{fig:speedup}
\end{figure}

Table~\ref{tab:controls} and Figure~\ref{fig:speedup} report the wall-time measurements that resolve the structural ambiguity. On the complete graph $K_n$, the reduced-observable run is faster than the full run by a factor that grows substantially with $N$, reaching approximately $8\times$ at $K_{16}$ ($T_{\mathrm{red}}=4.9 \pm 5.9$\,s versus $T_{\mathrm{full}}=38.5 \pm 15.1$\,s, $p=1$). On the star graph, the same pattern appears once $N$ crosses out of the overhead-dominated regime, with Star $N=16$ at $p=1$ showing $T_{\mathrm{red}}=2.5 \pm 2.2$\,s versus $T_{\mathrm{full}}=7.6 \pm 1.4$\,s, about $3\times$. Both families exhibit substantial wall-time savings despite having representative RCCs that span every qubit, which directly refutes the qubit-span hypothesis. The random 3-regular family at $N\geq 14$, where $|\mathcal{O}_E|=|E|$, shows no speedup ($T_{\mathrm{red}} \approx T_{\mathrm{full}}$). The qualitative pattern across all three controls is consistent with the prediction of Section~\ref{est-cost}, namely that the saving is set by the orbit-count reduction in the classical aggregation step.

\begin{table}[htbp]
\centering
\resizebox{\textwidth}{!}{
\begin{tabular}{|l|c|c|c|c|c|c|c|}
\hline
Graph & $N$ & $p$ & $|\mathcal{O}_E|$ & $T_{\mathrm{red}}$ (s) & $T_{\mathrm{full}}$ (s) & $R_{\mathrm{red}}$ & $R_{\mathrm{full}}$\\ [0.5ex]
\hline\hline
Complete $K_n$ & 8  & 1 & 1 & $1.68\pm0.24$ & $2.02\pm0.02$ & $1.000\pm0.000$ & $1.000\pm0.000$\\
\hline
Complete $K_n$ & 12 & 1 & 1 & $1.66\pm0.24$ & $3.96\pm1.02$ & $1.000\pm0.000$ & $0.991\pm0.013$\\
\hline
Complete $K_n$ & 14 & 1 & 1 & $2.04\pm0.25$ & $8.49\pm2.25$ & $0.993\pm0.010$ & $1.000\pm0.000$\\
\hline
Complete $K_n$ & 16 & 1 & 1 & $4.90\pm5.90$ & $38.5\pm15.08$ & $1.000\pm0.000$ & $1.000\pm0.000$\\
\hline
Complete $K_n$ & 16 & 2 & 1 & $49.6\pm5.29$ & $68.8\pm22.2$ & $0.953\pm0.066$ & $1.000\pm0.000$\\
\hline
Star, $n$ vertices & 8  & 1 & 1 & $1.68\pm0.16$ & $1.64\pm0.13$ & $1.000\pm0.000$ & $1.000\pm0.000$\\
\hline
Star, $n$ vertices & 12 & 1 & 1 & $1.40\pm0.21$ & $2.32\pm0.07$ & $1.000\pm0.000$ & $1.000\pm0.000$\\
\hline
Star, $n$ vertices & 14 & 1 & 1 & $1.50\pm0.30$ & $3.36\pm0.60$ & $0.974\pm0.036$ & $0.974\pm0.036$\\
\hline
Star, $n$ vertices & 16 & 1 & 1 & $2.54\pm2.18$ & $7.61\pm1.38$ & $0.933\pm0.054$ & $0.956\pm0.063$\\
\hline
Random 3-regular & 14 & 1 & 21 & $3.79\pm0.77$ & $3.75\pm0.65$ & $0.860\pm0.066$ & $0.842\pm0.074$\\
\hline
Random 3-regular & 14 & 2 & 21 & $5.52\pm0.88$ & $5.61\pm0.46$ & $0.754\pm0.050$ & $0.702\pm0.025$\\ [0.5ex]
\hline
\end{tabular}}
\caption{Diagnostic-control wall-time measurements (mean $\pm$ std over three runs), CPU backend. The complete graph $K_n$ and the star graph have a single edge orbit and representative RCCs that span every qubit. Both exhibit substantial wall-time savings that grow with $N$, refuting the qubit-span hypothesis. Random 3-regular graphs at $N=14$ have $|\mathcal{O}_E|=|E|$ and exhibit no saving. The pattern is consistent with the orbit-count prediction of Section~\ref{est-cost}.}
\label{tab:controls}
\end{table}

\subsubsection{Architecture independence: CPU versus GPU}
\begin{figure}[htbp]
\centering
\includegraphics[width=\linewidth]{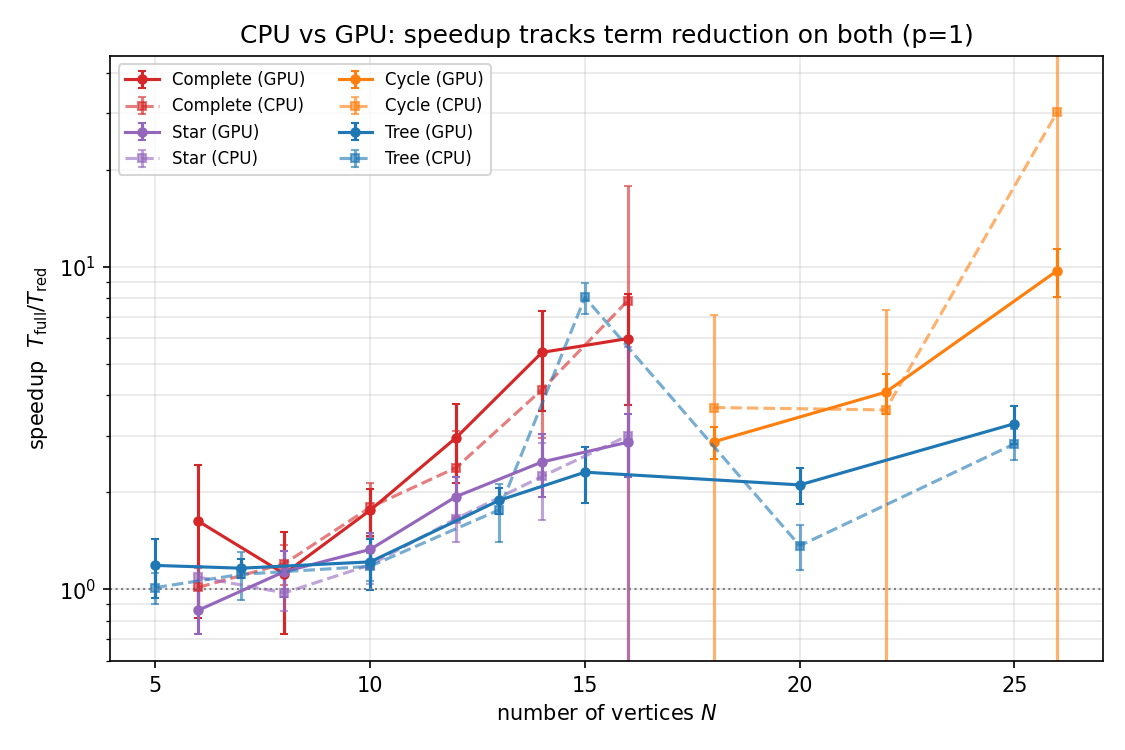}
\caption{(Color Online) CPU versus GPU wall-time saving on the diagnostic-control families. The two backends overlap within run-to-run variance for every family, confirming that the saving is from the classical aggregation step in Eq.~\ref{eq:agg} rather than from a backend-specific quantity.}
\label{fig:cpu_gpu}
\end{figure}

Figure~\ref{fig:cpu_gpu} reproduces the controlled study on a GPU backend (Qiskit Aer GPU statevector). The wall-time saving pattern is preserved across architectures, with the complete graph showing the same growth in saving with $N$ and the random 3-regular family showing no saving on either backend. CPU and GPU traces overlap within the run-to-run variance. The architecture independence is the third diagnostic, ruling out an architecture-specific cost as the source of the saving. The aggregation step in Eq.~\ref{eq:agg} is a classical-side post-processing of statevector amplitudes that is dispatched on the host CPU in both backends, which is why the orbit-count saving carries through.

\subsubsection{Approximation-ratio preservation}
\begin{figure}[htbp]
\centering
\includegraphics[width=\linewidth]{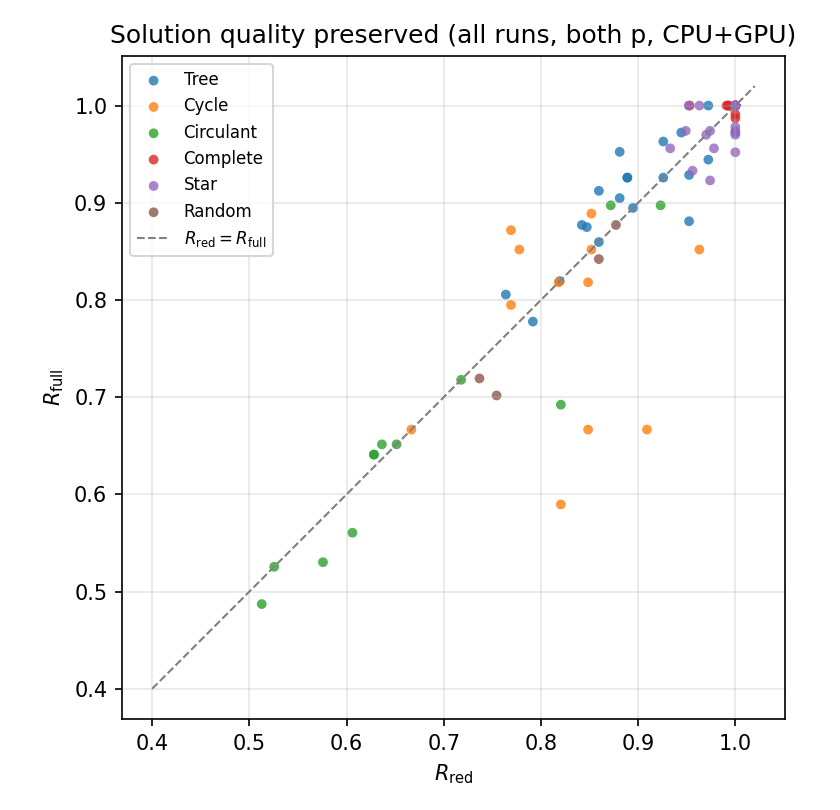}
\caption{(Color Online) Approximation ratios $R_{\mathrm{red}}$ (orange) and $R_{\mathrm{full}}$ (blue) across the diagnostic-control families. The two overlap within optimizer variance for every instance, consistent with $\langle H_{\mathrm{red}}\rangle = \langle H_C\rangle$ from Eq.~\ref{eq:exact}.}
\label{fig:quality}
\end{figure}

Figure~\ref{fig:quality} reports the approximation ratios $R_{\mathrm{red}}$ and $R_{\mathrm{full}}$ for every controlled instance. The two are equal within optimizer variance across all families and all $N$ tested, consistent with the algebraic identity in Eq.~\ref{eq:exact}. The reduced observable preserves solution quality exactly under sampling noise.

\subsubsection{Bounded-degree non-tree families}
Beyond the diagnostic controls, we report wall-time measurements on the bounded-degree non-tree families used to characterize the speedup magnitude on graphs more typical of the QAOA literature (Table~\ref{tab:nontree}). The cycle, generalized Petersen, and circulant families all exhibit speedups consistent with their orbit ratios, with the circulant $C_{13}(1,2)$ at $p=1$ showing $T_{\mathrm{red}}=1.44 \pm 0.14$\,s versus $T_{\mathrm{full}}=2.27 \pm 0.46$\,s. Random 3-regular at $n=12$, with $|\mathcal{O}_E|=6$ orbits, shows a small but non-zero saving consistent with that orbit count. Approximation ratios satisfy $R_{\mathrm{red}}\approx R_{\mathrm{full}}$ throughout. At these qubit counts the absolute wall times are small and partially overhead-influenced. The pronounced speedups on the larger tree instances of Section~\ref{RD} (the 31-qubit and 34-qubit cases) indicate the magnitude attained once the aggregation cost dominates the wall time.

\begin{table}[htbp]
\centering
\resizebox{\textwidth}{!}{
\begin{tabular}{|l|c|c|c|c|c|c|}
\hline
Graph & $p$ & $|\mathcal{O}_E|$ & $T_{\mathrm{red}}$ (s) & $T_{\mathrm{full}}$ (s) & $R_{\mathrm{red}}$ & $R_{\mathrm{full}}$\\ [0.5ex]
\hline\hline
Cycle $C_{13}$ & 1 & 1 & $1.66\pm0.79$ & $2.10\pm0.46$ & $1.000\pm0.000$ & $1.000\pm0.000$\\
\hline
Cycle $C_{13}$ & 2 & 1 & $3.14\pm0.98$ & $3.42\pm0.78$ & $1.000\pm0.000$ & $0.967\pm0.067$\\
\hline
Petersen $GP(5,2)$ & 1 & 1 & $1.86\pm0.44$ & $2.17\pm0.87$ & $0.983\pm0.033$ & $1.000\pm0.000$\\
\hline
Petersen $GP(5,2)$ & 2 & 1 & $2.05\pm0.37$ & $2.97\pm0.82$ & $0.917\pm0.129$ & $0.900\pm0.122$\\
\hline
Circulant $C_{13}(1,2)$ & 1 & 2 & $1.44\pm0.14$ & $2.27\pm0.46$ & $1.000\pm0.000$ & $0.978\pm0.044$\\
\hline
Circulant $C_{13}(1,2)$ & 2 & 2 & $2.43\pm0.45$ & $2.86\pm0.36$ & $0.911\pm0.083$ & $0.911\pm0.130$\\
\hline
Random 3-regular $n{=}12$ & 1 & 6 & $1.89\pm0.41$ & $1.97\pm0.26$ & $0.912\pm0.064$ & $0.875\pm0.068$\\
\hline
Random 3-regular $n{=}12$ & 2 & 6 & $2.51\pm0.27$ & $2.78\pm0.86$ & $0.812\pm0.143$ & $0.875\pm0.112$\\ [0.5ex]
\hline
\end{tabular}}
\caption{Wall-time measurements on bounded-degree non-tree families at $p=1$ and $p=2$, mean $\pm$ std over five runs. The cycle, Petersen, and circulant families show savings consistent with their orbit ratios. Approximation ratios satisfy $R_{\mathrm{red}}\approx R_{\mathrm{full}}$.}
\label{tab:nontree}
\end{table}

\subsection{Scope and limitations}\label{scope}
The empirical results reported in Sections~\ref{RD} and~\ref{diag} demonstrate that AA-QAOA accelerates the classical statevector simulation of QAOA on graphs with non-trivial automorphism group, with the saving set by the orbit-count reduction in the cost-expectation aggregation step. Three scope statements are worth making explicitly so that the contribution is not over-claimed.

First, the saving is a property of the classical estimator. On a quantum processor, the full QAOA circuit executes regardless of the measured observable, all diagonal ZZ terms of $H_C$ are co-measured at the cost of a single shot batch, and the aggregation reduces to a trivial pass over classical bitstrings. There is no AA-QAOA QPU speedup, by design and by mechanism. The relevant audience is the substantial body of QAOA research that runs entirely on classical simulators, where the aggregation step is a real cost.

Second, the saving is graph-dependent. Random 3-regular and other generically asymmetric graphs have trivial automorphism group at large $N$, with one orbit per edge, and AA-QAOA gives no saving on those instances. The method is operationally relevant only when the orbit ratio $|E|/|\mathcal{O}_E|$ is materially above one, which covers a substantial fraction of structured graphs studied in the QAOA literature, including bounded-degree edge-transitive families (cycles, circulants, hypercubes, generalized Petersen graphs), dense symmetric graphs (complete, complete bipartite), and tree families with regular branching.

Third, the orbit pre-computation has a cost not amortized in the per-iterate wall times reported here. Automorphism computation via Nauty is worst-case exponential in $|V|$, although for the families studied in this paper Nauty completes in seconds on a single CPU core. For larger or less-symmetric graphs the pre-computation could dominate, in which case AA-QAOA is appropriate only when the resulting orbit reduction is large enough to amortize the pre-computation across many optimizer iterates, which is the usual regime for QAOA simulation studies that sweep over $p$, initial conditions, or graph instances.

\section{Conclusion}\label{concl}
We have introduced AA-QAOA, an observable-substitution method that accelerates the classical statevector simulation of QAOA on graphs whose automorphism group is non-trivial. The substitution replaces the cost Hamiltonian $H_C$ with an orbit-reduced observable $H_{\mathrm{red}}$ that retains one representative ZZ term per edge orbit, weighted by orbit size. Because the prepared state is Aut$(G)$-invariant, the identity $\langle H_{\mathrm{red}}\rangle = \langle H_C\rangle$ (Eq.~\ref{eq:exact}) holds exactly, so optimization under $H_{\mathrm{red}}$ converges to the same parameters and the same approximation ratio as optimization under $H_C$. The wall-time saving derives from the classical aggregation step of the expectation evaluation, whose cost drops from $\mathcal{O}(|E|\cdot 2^n)$ to $\mathcal{O}(|\mathcal{O}_E|\cdot 2^n)$ per optimizer iterate.

The empirical study reports wall-time savings on tree-structured graphs up to 34 vertices, with the 34-qubit instance crossing from over 3600\,s under $H_C$ to 360\,s under $H_{\mathrm{red}}$, a reduction exceeding $90\%$, and a corresponding drop in peak memory consumption. Across six non-tree families spanning the symmetry spectrum, including the complete graph $K_n$, the star graph, the cycle, the generalized Petersen, the circulant, and random 3-regular instances, the diagnostic-control study isolates the orbit-count reduction $|E|/|\mathcal{O}_E|$ as the predictor of the saving. The complete graph $K_{16}$ at $p=1$ exhibits an approximately $8\times$ wall-time reduction despite spanning every qubit in its single representative RCC, directly refuting the qubit-span hypothesis. Random 3-regular graphs at $N\geq 14$ have $|\mathcal{O}_E|=|E|$ and show no saving, the expected negative control. A CPU-versus-GPU control reproduces the saving on both architectures, confirming that the mechanism is the classical aggregation step rather than a backend-specific quantity. Approximation ratios $R_{\mathrm{red}}\approx R_{\mathrm{full}}$ throughout, consistent with Eq.~\ref{eq:exact}.

The intended use case is the substantial body of QAOA research that operates entirely on classical statevector simulators, where the per-iterate aggregation cost is a real bottleneck on graphs with hundreds of edges. We are explicit that AA-QAOA does not transfer to a QPU run of the same circuit, on which the diagonal ZZ terms of $H_C$ are co-measured at the cost of a single shot batch and the aggregation is trivial. The method should be read as a classical-simulation accelerator that exposes structural symmetry to the estimator, not as a hardware-side optimization. Natural extensions include weighted MaxCut Hamiltonians, more general Ising-form objectives with non-trivial automorphism, and integration with the cone-subcircuit evaluation strategies developed in the recent QAOA-on-symmetric-graphs literature, where the orbit reduction and the locality reduction could compound.

\section*{Declarations}
\textbf{Funding:} The author received no funding for this work.\\
\textbf{Competing interests:} The author declares no competing interests.\\
\textbf{Data and code availability:} The graph-generation, automorphism-orbit, and QAOA benchmarking code, and the raw timing/approximation-ratio data underlying the tables and figures in this manuscript, are available from the author upon reasonable request.\\
\textbf{Author contribution:} V.N.P. conceived the method, designed and conducted all experiments, analyzed the results, and wrote and reviewed the manuscript.\\
\textbf{Use of AI:} The author acknowledges the use of Claude (Anthropic, Sonnet 5) in this work. See Section~\ref{AA-QAOA} for the specific disclosure. All AI-assisted analysis, derivations, and content were independently verified by the author, who takes full accountability for the manuscript's scientific claims.

\bibliographystyle{unsrt}
\bibliography{bibliography}

@article{adapt-qaoa,
  title = {Adaptive quantum approximate optimization algorithm for solving combinatorial problems on a quantum computer},
  author = {Zhu, Linghua and Tang, Ho Lun and Barron, George S. and Calderon-Vargas, F. A. and Mayhall, Nicholas J. and Barnes, Edwin and Economou, Sophia E.},
  journal = {Phys. Rev. Res.},
  volume = {4},
  issue = {3},
  pages = {033029},
  numpages = {9},
  year = {2022},
  month = {Jul},
  publisher = {American Physical Society},
  doi = {10.1103/PhysRevResearch.4.033029},
  url = {https://link.aps.org/doi/10.1103/PhysRevResearch.4.033029}
}

@article{greedy-qaoa,
  title = {Recursive greedy initialization of the quantum approximate optimization algorithm with guaranteed improvement},
  author = {Sack, Stefan H. and Medina, Raimel A. and Kueng, Richard and Serbyn, Maksym},
  journal = {Phys. Rev. A},
  volume = {107},
  issue = {6},
  pages = {062404},
  numpages = {14},
  year = {2023},
  month = {Jun},
  publisher = {American Physical Society},
  doi = {10.1103/PhysRevA.107.062404},
  url = {https://link.aps.org/doi/10.1103/PhysRevA.107.062404}
}

@article{shortcut-qaoa,
  title = {Shortcuts to the quantum approximate optimization algorithm},
  author = {Chai, Yahui and Han, Yong-Jian and Wu, Yu-Chun and Li, Ye and Dou, Menghan and Guo, Guo-Ping},
  journal = {Phys. Rev. A},
  volume = {105},
  issue = {4},
  pages = {042415},
  numpages = {10},
  year = {2022},
  month = {Apr},
  publisher = {American Physical Society},
  doi = {10.1103/PhysRevA.105.042415},
  url = {https://link.aps.org/doi/10.1103/PhysRevA.105.042415}
}

@article{warm-qaoa,
  doi = {10.22331/q-2023-09-26-1121},
  url = {https://doi.org/10.22331/q-2023-09-26-1121},
  title = {Warm-{S}tarted {QAOA} with {C}ustom {M}ixers {P}rovably {C}onverges and {C}omputationally {B}eats {G}oemans-{W}illiamson's {M}ax-{C}ut at {L}ow {C}ircuit {D}epths},
  author = {Tate, Reuben and Moondra, Jai and Gard, Bryan and Mohler, Greg and Gupta, Swati},
  journal = {{Quantum}},
  issn = {2521-327X},
  publisher = {{Verein zur F{\"{o}}rderung des Open Access Publizierens in den Quantenwissenschaften}},
  volume = {7},
  pages = {1121},
  month = sep,
  year = {2023}
}

@article{mfoa,
  title = {Mean-Field Approximate Optimization Algorithm},
  author = {Misra-Spieldenner, Aditi and Bode, Tim and Schuhmacher, Peter K. and Stollenwerk, Tobias and Bagrets, Dmitry and Wilhelm, Frank K.},
  journal = {PRX Quantum},
  volume = {4},
  issue = {3},
  pages = {030335},
  numpages = {19},
  year = {2023},
  month = {Sep},
  publisher = {American Physical Society},
  doi = {10.1103/PRXQuantum.4.030335},
  url = {https://link.aps.org/doi/10.1103/PRXQuantum.4.030335}
}

@article{fermionic-qaoa,
  title = {Fermionic quantum approximate optimization algorithm},
  author = {Yoshioka, Takuya and Sasada, Keita and Nakano, Yuichiro and Fujii, Keisuke},
  journal = {Phys. Rev. Res.},
  volume = {5},
  issue = {2},
  pages = {023071},
  numpages = {15},
  year = {2023},
  month = {May},
  publisher = {American Physical Society},
  doi = {10.1103/PhysRevResearch.5.023071},
  url = {https://link.aps.org/doi/10.1103/PhysRevResearch.5.023071}
}

@article{scale-qaoa,
  doi = {10.22331/q-2022-12-07-870},
  url = {https://doi.org/10.22331/q-2022-12-07-870},
  title = {Scaling of the quantum approximate optimization algorithm on superconducting qubit based hardware},
  author = {Weidenfeller, Johannes and Valor, Lucia C. and Gacon, Julien and Tornow, Caroline and Bello, Luciano and Woerner, Stefan and Egger, Daniel J.},
  journal = {{Quantum}},
  issn = {2521-327X},
  publisher = {{Verein zur F{\"{o}}rderung des Open Access Publizierens in den Quantenwissenschaften}},
  volume = {6},
  pages = {870},
  month = dec,
  year = {2022}
}

@article{qaoa2,
  title = {QAOA-in-QAOA: Solving Large-Scale MaxCut Problems on Small Quantum Machines},
  author = {Zhou, Zeqiao and Du, Yuxuan and Tian, Xinmei and Tao, Dacheng},
  journal = {Phys. Rev. Appl.},
  volume = {19},
  issue = {2},
  pages = {024027},
  numpages = {18},
  year = {2023},
  month = {Feb},
  publisher = {American Physical Society},
  doi = {10.1103/PhysRevApplied.19.024027},
  url = {https://link.aps.org/doi/10.1103/PhysRevApplied.19.024027}
}

@misc{far2014,
      title={A Quantum Approximate Optimization Algorithm}, 
      author={Edward Farhi and Jeffrey Goldstone and Sam Gutmann},
      archivePrefix={arXiv},
      year={2014},
      eprint={1411.4028},
      primaryClass={quant-ph},
      url={https://arxiv.org/abs/1411.4028}, 
}

@article{classic-sym,
    author ={Shaydulin Ruslan and Hadfield, Stuart and Hogg, Tad and Safro, Ilya} ,
    title = {Classical symmetries and the Quantum Approximate Optimization Algorithm},
    journal = {Quantum Information Processing},
    year = {2021},
    volume = {20},
    issue ={11},
    pages = {1573-1332}
}

@article{obstacle-sym,
  title = {Obstacles to Variational Quantum Optimization from Symmetry Protection},
  author = {Bravyi, Sergey and Kliesch, Alexander and Koenig, Robert and Tang, Eugene},
  journal = {Phys. Rev. Lett.},
  volume = {125},
  issue = {26},
  pages = {260505},
  numpages = {6},
  year = {2020},
  month = {Dec},
  publisher = {American Physical Society},
  doi = {10.1103/PhysRevLett.125.260505},
  url = {https://link.aps.org/doi/10.1103/PhysRevLett.125.260505}
}

@article{quantum-sym,
    author = {Joardar, Soumalya and Mandal, Arnab},
    title = {Quantum symmetry of graph C$^{*}$-algebras associated with connected graphs},
    journal = {Infinite Dimensional Analysis, Quantum Probability and Related Topics},
    volume = {21},
    number = {03},
    pages = {1850019},
    year = {2018},
    doi = {10.1142/S0219025718500194},

    URL = { https://doi.org/10.1142/S0219025718500194
    },
    eprint = { https://doi.org/10.1142/S0219025718500194
    }
}

@misc{small-sym,
      title={On the Effects of Small Graph Perturbations in the MaxCut Problem by QAOA}, 
      author={Leonardo Lavagna and Simone Piperno and Andrea Ceschini and Massimo Panella},
      archivePrefix={arXiv},
      year={2024},
      eprint={2408.15413},
      primaryClass={quant-ph},
      url={https://arxiv.org/abs/2408.15413}, 
}

@article{error-sym,
  doi = {10.22331/q-2021-09-21-548},
  url = {https://doi.org/10.22331/q-2021-09-21-548},
  title = {Quantum {E}rror {M}itigation using {S}ymmetry {E}xpansion},
  author = {Cai, Zhenyu},
  journal = {{Quantum}},
  issn = {2521-327X},
  publisher = {{Verein zur F{\"{o}}rderung des Open Access Publizierens in den Quantenwissenschaften}},
  volume = {5},
  pages = {548},
  month = sep,
  year = {2021}
}

@article{exploit-qaoa,
  author={Shaydulin, Ruslan and Wild, Stefan M.},
  journal={IEEE Transactions on Quantum Engineering}, 
  title={Exploiting Symmetry Reduces the Cost of Training QAOA}, 
  year={2021},
  volume={2},
  number={},
  pages={1-9},
  doi={10.1109/TQE.2021.3066275}}

@article{nauty,
title = {Practical graph isomorphism, II},
journal = {Journal of Symbolic Computation},
volume = {60},
pages = {94-112},
year = {2014},
issn = {0747-7171},
doi = {https://doi.org/10.1016/j.jsc.2013.09.003},
url = {https://www.sciencedirect.com/science/article/pii/S0747717113001193},
author = {Brendan D. McKay and Adolfo Piperno}
}

@article{vqerev,
    author = {Tilly,Jules and Chen,Hongxiang and Cao,Shuxiang and Picozzi,Dario and Setia,Kanav and Li,Ying and Grant,Edward and Wossnig,Leonard and Rungger,Ivan and Booth,George H. and Tennyson,Jonathan },
    title = {The Variational Quantum Eigensolver: A review of methods and best practices},
    journal = {Physics Reports},
    year = {2022},
    volume = {986},
    pages = {1-128},
    doi = {https://doi.org/10.1016/j.physrep.2022.08.003}
}
\end{document}